# Analysis of bottleneck motion using Voronoi diagrams.

**J. Liddle, A. Seyfried and B. Steffen**

Forschungszentrum Jülich GmbH, JSC
52425 Jülich, Germany
Corresponding author: j.liddle@fz-juelich.de

**Abstract** Standard definitions of the density exhibit large fluctuations when the size of the measurement area is comparable with the size of a pedestrian. An alternative measurement method exists where a personal space, calculated through the Voronoi diagram, is assigned to each pedestrian. In this contribution this method is applied to an experiment studying motion through a bottleneck and the reduced fluctuations demonstrated. The integrated density also permits examination on much smaller spatial scales than the standard definition, the insights into the pedestrian motion this provides are discussed.

## Introduction

The standard definition of the density in a measurement area $A$, of area $|A|$, containing $N$ pedestrians,

$$\rho_S = \frac{N}{|A|},$$

exhibits large fluctuations when the size of the measurement area is comparable with the size of a pedestrian. These fluctuations are typically of the same order as the density measurement itself. By averaging over time these fluctuations can be removed at the expense of reducing the time resolution of the measurements.

An alternative definition exists, the integrated density [1], where a density distribution (step function) is assigned to each pedestrian, this density distribution is calculated through the Voronoi diagram [2]. At a given time, $t_i$, we have at set of positions for each pedestrian $\{\vec{x}_1(t_i), \vec{x}_2(t_i), \cdots, \vec{x}_M(t_i)\}$. We compute the Voronoi diagram for these points obtaining a set of cells, $A_i$, for each pedestrian $i$. These cells can be thought of as the *personal space* belonging to each pedestrian. With these cells a density distribution can be defined for each pedestrian.

$$\rho_i(\vec{x}) = \begin{cases} \dfrac{1}{|A|} & \text{if } \vec{x} \in A_i \\ 0 & \text{otherwise} \end{cases}$$

and the density inside a measurement area, $|A|$, is defined as

$$\rho_V = \frac{\int p(\vec{x}) d\vec{x}}{|A|} \text{ where } p(\vec{x}) = \sum \rho_i(\vec{x}).$$

This method provides several advantages. The reduced fluctuations mean an instantaneous estimate of the density is possible and the presence of non-stationary states can be unambiguously detected, which is not possible with the standard method, see figure 1. The integrated density can also provide microscopic information about the density, which will inform the development of microscopic models.

Our experiment was performed in 2006 in the wardroom of the 'Bergische Kaserne Düsseldorf', with a test group of soldiers [3]. The experimental setup allowed the influence of the bottleneck width and length to be probed. In this proceeding the experiment with varying width (between 90 to 250 cm), b, is considered. Wider bottlenecks with more test persons were studied than in previous attempts [4-6].

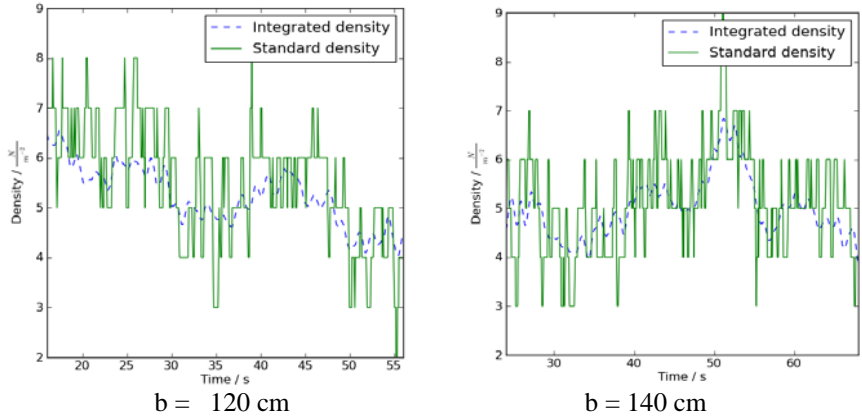

b = 120 cm        b = 140 cm

**Fig. 1. Density computed using standard and integral definitions. It can be clearly seen that the integrated definition fluctuates less than the standard definition. In the left figure the downward trend in the density is clearly visible when using the integrated density.**

In some of our experiments no steady state in the density was observed. The existence of the non-steady states can probably be attributed to the presence of highly motivated persons positioned at the front of the pack. The problems arising from trends in the density have not been discussed before, most likely due to insufficient resolution of the measurement method.

## Application

In figure 1 the density time series for two experiments are shown. It is immediately obvious that fluctuations are greatly reduced, and that an accurate estimate of the density can be obtained from a single frame.

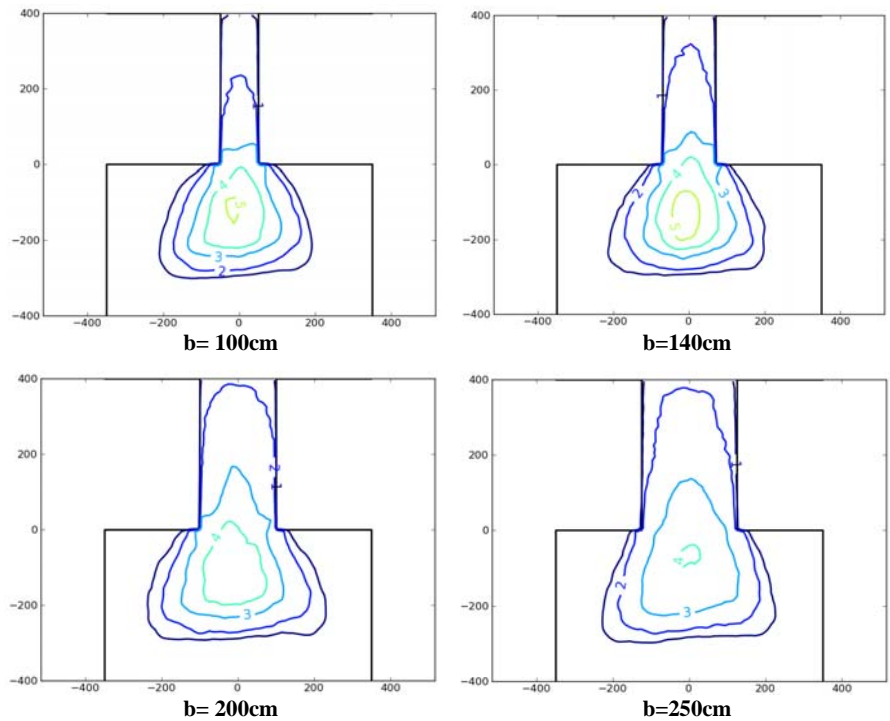

**Fig. 2. Density maps for bottlenecks of varying width. Average values during steady state.**

In addition to reduced fluctuations we can also obtain density measurements on small spatial scales, not obtainable with the standard definition. By calculating the integrated density over small (10 cm) regions and averaging over all configurations (during the steady state phase), maps of the density over the experimental area can be obtained. In figure 2 maps of the density are shown for a variety of widths, spanning the range of bottleneck widths studied. These maps reveal details of the density distribution like shape of congested areas. Firstly we note that the peak of the density lies around 125 cm from the bottleneck entrance, for all widths tested. As would be expected the peak of the density is sharpest in front of the narrowest bottlenecks. The shape of the high density region is also revealed, for the narrowest bottlenecks the high density region remains outside the bottleneck and for widest bottlenecks this region extends to the interior of the bottleneck.

## Conclusions

It has been shown that the integrated density exhibits greatly reduced fluctuations. Maps of the density over the experimental area can be obtained using the integrated method, these cannot be obtained, from available measurements, using the standard definition of the density and serve to further demonstrate the advantages of this method.

**Acknowledgments:** This study was supported by the Federal Ministry of Education and Research (BMBF), program on "Research for Civil Security - Protecting and Saving Human Life".